# Electrically controlled interlayer trion fluid in electron-hole bilayers

Ruishi Qi[1,2], Qize Li[1], Zuocheng Zhang[1], Sudi Chen[1,2], Jingxu Xie[1,2], Yunbo Ou[3], Zhiyuan Cui[1,2], David D. Dai[4], Andrew Y. Joe[5], Takashi Taniguchi[6], Kenji Watanabe[7], Sefaattin Tongay[3], Alex Zettl[1,2,8], Liang Fu[4], and Feng Wang[1,2,8,*]

[1] Department of Physics, University of California, Berkeley, CA 94720, USA

[2] Materials Sciences Division, Lawrence Berkeley National Laboratory, Berkeley, CA 94720, USA

[3] School for Engineering of Matter, Transport and Energy, Arizona State University, Tempe, AZ 85287, USA

[4] Department of Physics, Massachusetts Institute of Technology, Cambridge, MA 02139, USA

[5] Department of Physics, University of California, Riverside, CA 92521, USA

[6] International Center for Materials Nanoarchitectonics, National Institute for Materials Science, 1-1 Namiki, Tsukuba 305-0044, Japan

[7] Research Center for Functional Materials, National Institute for Materials Science, 1-1 Namiki, Tsukuba 305-0044, Japan

[8] Kavli Energy NanoSciences Institute, University of California Berkeley and Lawrence Berkeley National Laboratory, Berkeley, CA 94720, USA.

* To whom correspondence should be addressed: fengwang76@berkeley.edu

**Abstract:** The combination of repulsive and attractive Coulomb interactions in a quantum electron(e)-hole(h) fluid can give rise to novel correlated phases of multiparticle charge complexes such as excitons, trions and biexcitons. Here we report the first experimental realization of an electrically controlled interlayer trion fluid in two-dimensional van der Waals heterostructures. We demonstrate that in the strong coupling regime of electron-hole bilayers, electrons and holes in separate layers can spontaneously form three-particle trion bound states that resemble positronium ions in high energy physics. The interlayer trions can assume 1e-2h and 2e-1h configurations, where electrons and holes are confined in different transition metal dichalcogenide layers. We show that the two correlated holes in 1e-2h trions form a spin-singlet state with a spin gap of ~1meV. By electrostatic gating, the equilibrium state of our system can be continuously tuned into an exciton fluid, a trion fluid, an exciton-trion mixture, a trion-charge mixture or an electron-hole plasma. Upon optical excitation, the system can host novel high-order multiparticle charge complexes including interlayer four-particle complex (tetrons) and five-particle complex (pentons). Our work demonstrates a unique platform to study novel correlated phases of tunable Bose-Fermi mixtures and opens up new opportunities to realize artificial ions/molecules in electronic devices.

## Main

In many-body systems composed of oppositely charged particles, the combination of repulsive and attractive Coulomb interactions can stabilize a wide variety of multiparticle bound states. For example, interacting electrons and nuclei lead to the formation of atoms and molecules, and interacting electrons and positrons can generate two-particle positronium and three-particle positronium ions. In semiconductors, interacting electrons and holes can form excitons, trions, and biexcitons[1]. These multiparticle electron-hole (e-h) states in semiconductors are usually generated transiently through optical excitation, which makes it difficult to explore their steady-state quantum phases due to their short lifetime[2–4]. Van der Waals heterostructures composed of two transition metal dichalcogenide (TMD) semiconductor layers separated by a thin hexagonal boron nitride (hBN) layer enable the creation of electrically controlled strongly coupled e-h bilayers, which opens the opportunity to explore multiparticle e-h complexes at thermal equilibrium[5–11]. Previously, electrically controlled interlayer excitons (i.e., correlated e-h bound pairs) have been realized in MoSe$_2$/hBN/WSe$_2$ heterostructures, which lead to the observation of interlayer excitonic insulators (EI)[6,7] and perfect Coulomb drag behavior[12,13]. It is expected that such exciton fluid could lead to high-temperature exciton condensates[8,14]. Three- and four-particle e-h complexes have also been theoretically predicted to exist in e-h bilayers in the strong coupling limit, which could host completely new types of quantum behaviors[10,15]. Their experimental realization, however, has remained elusive so far.

In this article, we report the first experimental observation of a stable quantum fluid of interlayer trions, the three-particle e-h complexes, in an e-h bilayer. We experimentally demonstrate that when strongly coupled e-h fluids are electrostatically tuned to a commensurate e-h density ratio 2:1 or 1:2, strong Coulomb attraction together with quantum exchange correlation leads to the spontaneous creation of negative or positive trions – two electrons bound to one hole or two holes bound to one electron – in thermal equilibrium. Direct evidence of trion formation is revealed by optical spectroscopy, where the spin configuration of the trions strongly alters the allowed optical absorption process. We show that the two like charges in the trion form a spin-singlet with a spin gap of ~1 meV. The e-h bilayer system can be continuously tuned to host a 2D electron gas, a 2D hole gas, an interlayer exciton fluid, a positive trion fluid, a negative trion fluid, or a mixture of them. This opens the door to explore correlated Bose-Fermi mixtures with a rich phase diagram that includes the interplay among trion crystals, unpaired carriers and exciton superfluids[15]. Starting from the interlayer exciton and trion fluids in the e-h bilayer, we can further generate novel interlayer four-particle complex (tetrons) and five-particle complex (pentons) transiently with optical excitation.

## Strongly coupled electron-hole fluids

Figure 1a schematically shows the e-h bilayer device. A monolayer MoS$_2$ and a monolayer WSe$_2$ are separated by a monolayer or bilayer hBN tunneling barrier. Due to the type-II band alignment, the electrons and holes in the system naturally reside in MoS$_2$ and WSe$_2$ layers respectively. Such an e-h bilayer structure provides a highly tunable platform to control and study a strongly correlated Bose-Fermi mixture. The application of an interlayer bias voltage simply tunes the band alignment of the two layers[10,16]. When the single-particle band gap is electrically tuned below the interlayer exciton binding energy, the ground state of the system spontaneously forms e-h pairs that bind into interlayer excitons, which is arguably the first unambiguous realization of an EI state at zero magnetic field[6,7,12,13]. When the EI is doped with additional electrons or holes, the interlayer excitons can in principle bind with additional unpaired charges to create interlayer trions. However, it is more challenging to achieve spontaneous formation of interlayer trions in thermal equilibrium because they only appear in the very strong e-h coupling limit due to their small binding energy[17]. Here we fabricate e-h bilayer devices with significantly reduced interlayer distance (within 1nm) compared with previous studies. This enhances the e-h attraction relative to the electron-electron and hole-hole repulsion, favoring trion formation. With large e-h momentum mismatch suppressing the interlayer recombination and specially designed carrier reservoir ensuring low contact resistance (Methods and Extended Data Fig. 1), the e-h fluids remain near-equilibrium with the electrodes. The heterostructure is encapsulated by dielectric hBN on both sides and gated by few-layer graphene top gate (TG) and back gate (BG). We keep the electron layer grounded ($V_e = 0$) and apply voltages on the gates and the hole layer ($V_h$). The symmetric gate voltage $V_G \equiv V_{TG} + V_{BG}$ tunes the Fermi-level and thus the net charge density (e-h imbalance). With the antisymmetric gating $V_{TG} - V_{BG}$ fixed through our experiment, the bias voltage $V_B \equiv V_h - V_e$ tunes the band alignment. Fig. 1b is an optical image of such a device (D1) with a monolayer hBN spacer. While we focus on this device in the main text, our conclusions are also supported by highly reproducible data taken from another three devices (D2, D3 and D4) as summarized in Extended Data Figs. 2-4. The sample is kept at a temperature $T = 4.9$K unless otherwise specified.

The optical absorption spectrum of TMDs is sensitive to the charge density due to strong excitonic resonances and strong Coulomb interactions[18,19]. Taking advantage of this dependence, we determine the electron and hole densities as a function of applied voltages using the spectroscopy technique described in ref. [7]. Figure 1c shows the charge doping phase diagram as a function of $V_G$ and $V_B$, where red and green channels of the false color map encode the density of electrons ($n_e$) and holes ($n_h$) respectively. Black dashed lines approximately trace the commensurate density ratio between $n_h$ and $n_e$. Fig. 1d-e shows two representative gate-dependent reflectivity scans at constant bias voltage $V_B = 0.68$V and 0.735V respectively. The energy-derivative of reflectivity, $dR/dE$, is plotted for better visibility. At low bias voltages such as the one shown in Fig. 1d, the system has a finite type-II band gap, so only one type of charge can enter the system at a time. Well-defined intralayer exciton peaks ($X_0$) for MoS$_2$ and WSe$_2$ are observed when both layers are undoped. Upon electron or hole doping, another absorption peak emerges at lower energy on the MoS$_2$ or WSe$_2$ side, which is commonly known as the intralayer trion peak or the charged exciton peak ($X^-$/$X^+$)[18,19]. When the bias voltage exceeds the gap energy at $V_B > 0.72$V, the ground state of the system contains a fluid of electrons and holes at the same time, as signified by the coexistence of the WSe$_2$ $X^+$ peak and MoS$_2$ $X^-$ peak in the reflection spectrum in Fig. 1e. In this regime, strong Coulomb attraction between the electrons and holes can lead to spontaneous formation of correlated bound states in thermal equilibrium. The e-h fluids can consist of possibly interlayer excitons, interlayer trions, unpaired electrons/holes, or a mixture of them.

We pay special attention to three states with commensurate density ratios $n_e:n_h = 1:1$, 2:1 and 1:2. At $n_e:n_h = 1:1$, electrostatically injected electrons and holes are known to pair into interlayer excitons and form a strongly correlated EI[6,7]. At $n_e:n_h = 1:2$, we note that the WSe$_2$ $X^+$ resonance exhibits a prominent feature (Figure 1e). Compared with nearby regions, the WSe$_2$ $X^+$ peak almost disappears at this density ratio, and a new peak (labeled as $P^+$) emerges at a slightly higher energy. A similar feature, although less obvious, is observed on the MoS$_2$ side at the carrier ratio $n_e:n_h = 2:1$. Although the spectral features are relatively weak, they are very reliable: the spectral change is much stronger than experimental uncertainties and is highly reproducible over multiple samples. The emerging peak centered at $n_e:n_h=1:2$ and 2:1 is the first indication of the formation of positive and negative interlayer trions, respectively. Because the WSe$_2$ $X^+$ peak is sharper and stronger, we will focus our analysis on the WSe$_2$ resonances. Figure 1f shows the $dR/dE$ signal at the WSe$_2$ $X^+$ peak energy over the entire e-h density phase space. We can observe a clear feature associated with the decreased signal at the WSe$_2$ $X^+$ resonance along the 1:2 ratio line.

## Spectroscopic signature of trions

Next, we confirm that the emergent $P^+$ peak is due to the formation of interlayer trions. It is well known that hydrogen anions – two electrons bound to one proton – only have one bound state with two electrons forming a spin-singlet. Trions are composed of strongly correlated three particle e-h complex that resembles hydrogen anions. A definitive signature of the trion is its internal spin structure: the two same-charge particles must form a spin-singlet pair to lower the energy in the bound trion state[15,20]. The magnetic-field dependence of the optical transitions therefore can provide an unambiguous spectroscopic signature of the interlayer trions.

Spin-valley locking combined with the optical selection rules in semiconducting TMDs allow us to selectively access the holes of particular spin using a circularly polarized laser probe[21–23]. Figs. 2a-b compare the magnetic field ($B$) dependence of helicity-resolved $dR/dE$ spectra at a fixed hole density $n_h = 1.0 \times 10^{12} \text{cm}^{-2}$ but different electron doping $n_e = 0$ and $n_e = n_h/2$ ($n_e:n_h$=1:2). The spectra are taken with left-handed circularly polarized light that selectively probes the spin-up component. When $n_e = 0$, the holes form a Fermi liquid, and the spin can be continuously polarized by the external magnetic field. The $X^+$ resonance intensity shows a monotonic increase with the applied magnetic field.

In contrast, distinct field-dependence is observed when $n_e = n_h/2$ (Fig. 2b). We observe two peaks with an energy separation of ~5 meV, and these two spectral features exhibit completely different magnetic field dependence. The $X^+$ peak (marked by the yellow dashed line) still shows an increase in intensity with the magnetic field, similar to the $n_e = 0$ case. However, its intensity is significantly lower. The new peak $P^+$ (marked by the purple dashed line) does not show a monotonic change with $B$; instead, it is the strongest at zero field, and decays symmetrically at higher field regardless of the field direction (Fig. 2b, inset). This unusual magnetic field dependence of the $P^+$ arises from the spin-singlet configuration of the two holes in the trion: the spin singlet cannot be continuously polarized. Instead, spin-singlet trions dissociate at high field when the Zeeman energy becomes comparable to the trion binding energy.

The resonant photon absorption process at zero magnetic field is schematically illustrated in Fig. 2c. When only free holes are initially present ($n_e = 0$), the intralayer trion resonance ($X^+$ peak) can be excited when the free hole and the photoexcited hole have opposite spin (and valley). At the density ratio $n_e = n_h/2$ the system forms positive interlayer trions in thermal equilibrium, where there are already two identical holes with singlet pairing bound to one electron. The optically generated e-h pair will interact with them in a distinctly different way. The five-particle state here involves three identical holes in the WSe$_2$ layer, which cannot all occupy the lowest orbit because the spin degeneracy can only accommodate two. The third one has to occupy a higher orbit, increasing the total energy. This leads to the $P^+$ peak at a higher energy.

Figure 2d and 2e compare the zero-field optical spectra at $n_e = 0$ and $n_e = n_h/2$. At $n_e = 0$, the system contains a well-understood 2D hole gas in WSe$_2$, where two absorption peaks, the intralayer exciton peak $X_0$ and the intralayer trion peak $X^+$, are observed. Both of them can be well described by a Fano line shape. At $n_e = n_h/2$, the spectrum can be described by three Fano peaks. The $X^+$ peak loses its oscillator strength significantly, and the high-energy $P^+$ peak appears. The

$X^+$ peak does not completely disappear at $n_e = n_h/2$, suggesting a portion of the interlayer trions are ionized due to the finite temperature and small trion binding energy.

The magnetic field dependence of the fitted oscillator strength allows us to estimate the interlayer trion binding energy $\epsilon_t$. At zero magnetic field, the ground state of the system is the spin-singlet trion, and excited states include a continuum of ionized states (unbound hole-exciton pairs) that are spin degenerate. When the field is weak, most holes are in the trion bound state despite a small portion of thermally ionized holes. With a strong external field, the Zeeman effect splits the energy levels of the spin-up and spin-down ionized states, but the trion state energy is not affected. As illustrated in the inset of Fig. 2f, this reduces the gap between the ground state and the continuum, ionizing a larger portion of the trions. Assuming that the oscillator strength of each optical transition peak depends linearly on the density of the corresponding particle, the spin-up and spin-down ionized hole density will be proportional to the $X^+$ oscillator strength at $B$ and $-B$ respectively, and the density of holes in the trion bound state will be proportional to the $P^+$ peak strength. The oscillator strength ratio from the Fano peak fitting is shown in Fig. 2f. With an effective g-factor of $g = 6.1$ (ref.[24]), 1T field amounts to 0.35meV Zeeman shift. The field required to ionize half of the trions is roughly 4T, indicating a trion binding energy on the order of 1 meV. We fit the experimental data to a simple model (Methods) and estimate the spin gap, or the trion binding energy, to be $\epsilon_t = 1.1 \pm 0.3$ meV. It is 1-2 orders of magnitude smaller than the interlayer exciton binding energy ($\epsilon_x = 42 \pm 5$ meV, Extended Data Fig. 5)[6,7]. This binding energy ratio is similar to the hydrogen anion compared to the hydrogen atom (0.75eV versus 13.6eV), and is also consistent with previous quantum Monte Carlo simulation results[17]. The temperature dependence of the trion occupation shown in Extended Data Fig. 6 also suggests a binding energy of ~1 meV, where half of the trions ionize at around $T \approx 10$ K.

## Electrically tunable Bose-Fermi mixtures

Since we can control the e-h density ratio by electrostatic gating, the ground state of the system can be electrically tuned to form different mixtures of excitons, trions and unpaired charges. Fig. 3a shows the WSe$_2$ resonances as a function of $n_e$ for a fixed hole doping density $n_h = 1.0 \times 10^{12} \text{cm}^{-2}$. The spectrum remains almost the same for $n_e > n_h$ as well as $n_e = 0$, but shows a significant change between $0 < n_e < n_h$ centered at $n_e = n_h/2$. Fig. 3b shows the fitted oscillator strength of the $X^+$ peak and the $P^+$ peak as a function of $n_e$. Starting from the 2D hole gas at $n_e = 0$, adding electrons into the system creates interlayer trions, decreasing the number of free holes. This leads to the initial drop in $X^+$ intensity and increase in $P^+$ intensity. The density ratio $n_e = n_h/2$ is most favorable for interlayer trions. The $X^+$ oscillator strength decreases to ~20% at this ratio, indicating that ~80% of the holes are in the positive trion state at our base temperature $T = 4.9$K, which amounts to a positive trion density $n_t^+ \approx 0.4 \times 10^{12} \text{cm}^{-2}$. When $n_e$ is further increased, the additional electrons will dissociate some interlayer trions since the exciton binding energy is larger than the trion binding energy, leading to a mixture of interlayer excitons and positive trions. At net charge neutrality $n_e = n_h$, the system turns into an EI phase (Extended Data Fig. 5). Further increasing the electron density makes negative trions favorable. Negative trions are composed of two electrons but only one hole, so they do not alter the WSe$_2$ absorption spectrum significantly because the spin degeneracy of 2 is sufficient to accommodate one additional photoexcited hole. Signatures of negative trions are manifested in a high energy peak ($P^-$) in the MoS$_2$ spectrum (Extended Data Fig. 7).

A 2D map of the experimentally estimated positive trion density as a function of $n_e$ and $n_h$ is displayed in Fig. 3c. To understand this, we consider a simple model at zero temperature. Starting from $n_e$ electrons and $n_h$ holes, they will first pair into excitons to lower the total energy. The exciton density would be min $(n_e, n_h)$, limited by the minority carrier. Then the min $(n_e, n_h)$ excitons and $|n_h - n_e|$ unpaired charges will further bind into a maximum number of trions, which is again limited by the minority of them. Thus, the trion density in this simple picture is given by $n_t = \min(n_e, n_h, |n_e - n_h|)$. There will be $n_x = \min(n_e, n_h) - n_t$ interlayer excitons and $n_u = |n_e - n_h| - n_t$ unpaired charges. Fig. 3d shows the interlayer trion (including positive and negative trions) density as a function of $n_e$ and $n_h$ according to this model. The trion density is zero along the diagonal line (interlayer exciton fluid), the horizontal axis (2D electron gas) and the vertical axis (2D hole gas). The two peaks along the $n_e : n_h = 1 : 2$ line and the 2:1 line correspond to positive and negative trions respectively. The positive trion sector agrees nicely with the experimental estimation using the WSe$_2$ spectra, while the negative trion sector is also similar to the MoS$_2$ results shown in Extended Data Fig. 7. Apart from these commensurate ratios, the ground state of the e-h bilayer can be continuously tuned to be a hole-trion mixture ($n_e < n_h/2$), an exciton-trion mixture ($n_h/2 < n_e < 2n_h$), or an electron-trion mixture ($n_e > 2n_h$), as illustrated in Fig. 3e.

## High order multiparticle complexes

The exciton fluid and the trion fluid can also host novel high order multiparticle states upon photoexcitation. We have already seen the prominent P$^+$ peak at $n_e = n_h/2$ as an example, corresponding to five-particle bound complexes created by photoexcited intralayer e-h pairs interacting with positive interlayer trions. For the case of $n_e = 0$, $n_e = n_h$ and $n_e = 2n_h$, we only observe a X$^+$ peak and the reflection spectra are very similar among them. However, there is actually a small spectral difference at different density ratios (Figs. 4a-c). Their resonance energy, together with the interlayer exciton binding energy $\epsilon_x = 42 \pm 5$ meV and trion binding energy $\epsilon_t = 1.1 \pm 0.3$ meV (Fig. 4d and Methods), allows us to determine the relative energy levels of 3-particle, 4-particle and 5-particle bound states (Fig. 4e-h).

We start with the simple 3-particle case (Figs. 4a, 4e). The X$^+$ peak can be well fitted by a Fano resonance. Subtracting the X$_0$ peak energy in the low doping limit (1.727 eV) from the fitted energy gives the intralayer trion binding energy of $\epsilon_3 = 20.6$ meV.

Figs. 4a-b compares the X$^+$ peak at the $n_e = 0$ state (2D hole gas) and the $n_e = n_h$ state (dipolar EI). There is a very small but reproducible energy shift of $0.2 \pm 0.1$ meV for the monolayer hBN spacer device and $0.4 \pm 0.1$ meV for bilayer hBN spacer devices (Extended Data Figs. 2-4). This suggests the formation of optically generated tetron state (sometimes referred to as a Bose polaron as opposed to the Fermi polaron)[25] when the system has an interlayer exciton fluid ground state. As schematically illustrated in Fig. 4f, the tetron can be viewed as an interlayer-intralayer hybrid biexciton state. The fitted peak energy is $\epsilon_4 = 20.8$ meV lower than the X$_0$ peak, indicating a very similar binding energy to the intralayer trion illustrated in Fig. 4e. The energy difference being small is consistent with a recent theoretical prediction[25]. Despite the small spectral change, they correspond to distinct objects.

Comparison between the resonance energy in the 2D hole gas ($n_e = 0$, Fig. 4a) and in the negative trion fluid ($n_e = 2n_h$, Fig. 4c) shows a vanishingly small energy difference. It indicates that the five-particle bound state (two MoS$_2$ electrons, one WSe$_2$ electron and two WSe$_2$ holes, as illustrated in Fig. 4g) has almost the same binding energy of 20.6 meV as the WSe$_2$ intralayer trion. This behavior is very different from the high-energy P$^+$ peak discussed earlier in the positive trion case. We note that the five-particle state associated with the negative interlayer trion does not involve three identical particles: two MoS$_2$ electrons and two WSe$_2$ holes can both form spin-singlets and stay in the lowest orbit. Compared with the WSe$_2$ intralayer trion state, the absorption resonance energy is minimally affected by the presence of electrons in the opposite layer.

Finally, Fig. 4h illustrates the excited states of the positive trion fluid after photoexcitation, or the P$^+$ peak discussed earlier. The binding energy of the penton – an electron in MoS$_2$, an electron in WSe$_2$, and three holes in WSe$_2$ all bound together – can be determined from the fitting in Fig. 2e. It is 15.5 meV lower than the energy of an interlayer trion far separated from an intralayer exciton. However, this is not the lowest-energy state of these five particles. Because the interlayer trion binding energy is very small, an intralayer exciton can absorb one hole from an interlayer trion, forming an intralayer trion and an interlayer exciton. This process lowers the energy by $\epsilon_3 - \epsilon_t = 19.5 \text{ meV} > 15.5 \text{ meV}$. Therefore, the optically excited penton is unstable and will quickly dissociate into two parts after the photoexcitation.

## Summary and outlook

We have reported the first experimental realization of trion fluid in thermal equilibrium. This trion fluid is promising for many exotic quantum phases. Their charged nature leads to strong inter-trion Coulomb repulsion, and their heavy mass suppresses the kinetic energy. The dominance of Coulomb energy over kinetic energy can drive the system to spontaneously break the continuous translation symmetry and form a quantum crystal of trions[15,26,27]. Recent theories also predicted the possibility of interlayer trion mediated topological superconductivity[28]. In addition, we demonstrated novel Bose-Fermi mixtures composed of tunable mixture of interlayer trions, interlayer excitons and unpaired charges. As a condensed matter analogue of two-dimensional molecules, they provide new opportunities to explore quantum physics and chemistry with electrical tunability in the strongly correlated regime.

## Methods

**Device fabrication.** We have fabricated and measured multiple e-h bilayer devices (D1-D4). The thin hBN spacer can be either a monolayer (D1) or a bilayer (D2-D4), and the electron layer can be either MoS$_2$ (D1, D4) or MoSe$_2$ (D2, D3). All these devices give consistent results.

High-quality MoS$_2$ crystals are grown by chemical vapor transport[29,30] while MoSe$_2$ and WSe$_2$ crystals are grown by the flux method. Monolayer MoS$_2$, monolayer MoSe$_2$, monolayer WSe$_2$, few-layer graphene and hBN flakes are mechanically exfoliated from bulk crystals onto Si/SiO$_2$ substrates. We use a dry-transfer method based on polyethylene terephthalate glycol (PETG) stamps to fabricate the heterostructures. A 0.5 mm thick clear PETG stamp is employed to sequentially pick up the flakes at 65-75 °C. The whole stack is then released onto a high resistivity Si substrate with a 90 nm SiO$_2$ layer at 95-100 °C, followed by dissolving the PETG in chloroform at room temperature for one day. Electrodes (50-70 nm Au with 5 nm Cr adhesion layer) are defined using photolithography (Durham Magneto Optics, MicroWriter) and electron beam evaporation (Angstrom Engineering).

In order to close the type-II band gap of 1.15eV (1.5eV) between $MoS_2/WSe_2$ ($MoSe_2/WSe_2$) by voltage, a large interlayer bias needs to be applied. Because the hBN tunneling barrier is not perfectly opaque, a finite interlayer tunneling current exists. To reduce the interlayer leakage, we apply a large vertical electric field $V_{TG} - V_{BG}$ to assist closing the gap. 5-8 nm thick hBN is used as the gate dielectric because it can sustain a large electric field ($> 0.5$V/nm) without breaking down. The $MoS_2$ ($MoSe_2$) layer and $WSe_2$ layer are deliberately angle-misaligned to create a large momentum mismatch and suppress interlayer tunneling. The interlayer leakage current shown in Extended Data Fig. 8 gives a lower bound of the interlayer recombination lifetime $\tau$. We assume all the interlayer leakage is from the region of interest, and the lifetime can be estimated as $\tau = en_x A/I$, where $A$ is the heterostructure area, $n_x = \min(n_e, n_h)$ is the e-h pair density, and $I$ is the leakage current. The recombination lifetime is estimated to be ~1 microsecond. The estimated lifetime differs by a factor of ~10 among four devices, presumably due to different hBN spacer thicknesses, different type-II bandgaps, and different angle mismatches between two TMD layers. The behavior of the interlayer trions remain almost identical among different devices, nevertheless. The lifetime is 6 orders of magnitudes longer than the optically excited intralayer excitons and trions with typical lifetimes on the order of a picosecond[2].

The finite interlayer leakage current requires good electrical contacts to the TMD layers. Otherwise, the voltage will be mostly dropped on the contact and the bias cannot be efficiently applied. We improve the contacts to TMD layers with specially designed contact regions (Extended Data Fig. 1). We misalign the top and back gates with the heterostructure such that there is a $MoS_2$ ($MoSe_2$) contact region ($n^{--}$) that is only covered by the top gate while the $WSe_2$ contact region ($p^{++}$) is only gated by the back gate. This makes the contact region very heavily doped ($> 10^{13} \text{cm}^{-2}$) and the contact resistance at the graphite-TMD junction is reduced, ensuring efficient charge injection into the heterostructure region. However, the injection of interlayer excitons and trions is a correlated injection process that requires wavefunction overlap between electrons and holes[6]. Therefore, we further improve the contact by inserting a 3-5nm hBN spacer between the TMD layers near the contact region. Compared with the region of interest, the vertical electric field $V_{TG} - V_{BG}$ creates a larger voltage difference due to the increased interlayer distance in this region, turning the heterostructure into type-III band alignment with a large band overlap. This region serves as a carrier reservoir and ensures efficient charge, exciton and trion injection into the region of interest.

**Optical measurements.** The optical measurements are performed in an optical cryostat (Quantum Design, OptiCool) with a nominal base temperature down to $T = 2$ K. The actual sample temperature $T = 4.9$K was calibrated by a thermometer placed at the sample position. The gate and bias voltages are applied by Keithley 2400 or 2450 source meters while the leakage current is monitored. The reflection spectroscopy is performed with a supercontinuum laser (Fianium Femtopower 1060 Supercontinuum Laser) as the light source. The laser is focused on the sample by a 20× Mitutoyo objective with ~1.5µm beam diameter. A small beam size provides a local probe that we can park in a clean region free of bubbles. We choose a low incident laser power (30-60 nW) to minimize photodoping effects. The reflected light is collected by a spectrometer (Princeton Instruments PIXIS 256e) with 1000 ms exposure time. To minimize the influence of laser fluctuations, another laser beam directly reflected from a silver mirror is simultaneously collected to normalize the sample reflection spectra.

**Determination of trion and exciton binding energy.** We fit the experimental magnetic field dependence data to the following model to estimate the interlayer trion binding energy. Consider the case of $n_h$ holes and $n_e = n_h/2$ electrons. At low but finite temperatures (a few Kelvin), we assume the interlayer excitons are always bound because the exciton binding energy is significantly larger than the thermal energy or the magnetic field energy scale. The system can either form trions or have exciton-hole unbound pairs. The trion bound state has an energy level of $-\epsilon_t$ regardless of the magnetic field. The continuum of unbound holes starts from $\pm g\mu_B B$ for the two spins respectively, where $\mu_B$ is the Bohr magneton. Then the total density of states for the hole is $D(E) = n_h \delta(E + \epsilon_t) + D_0 \Theta(E + g\mu_B B) + D_0 \Theta(E - g\mu_B B)$. Here $D_0$ is the effective density of states for the parabolic bands of one particular spin, and $\delta$ and $\Theta$ are the Dirac delta function and the Heaviside step function. Then the number of trions and free holes can be calculated by integrating the corresponding part of the density of states multiplied by the Fermi-Dirac distribution function. The experimental data is fitted to such a model with $\epsilon_t$ and $D_0$ as fitting parameters.

The interlayer exciton binding energy is determined by the charge gap of the EI phase in the limit of zero exciton density. The charge compressibility $\partial(n_e - n_h)/\partial V_G$ shown in Extended Data Fig. 5 reveals the exciton-compressible, charge-incompressible EI phase[6,7]. The width at the bottom of the triangular region is $2\epsilon_x \approx 84$meV. The determined interlayer exciton binding energy is $\epsilon_x = 42 \pm 5$meV for the monolayer hBN spacer device and $37 \pm 4$meV for bilayer hBN spacer devices.

## Data availability
The data that support the findings of this study are available from the corresponding author upon request.

## Acknowledgements
We thank Dr. Ivan Amelio, Dr. Cedric Robert, and Dr. Xavier Marie for their insightful discussion. The magneto-optic spectroscopy and data analysis were supported by the U.S. Department of Energy, Office of Science, Office of Basic Energy Sciences, Materials Sciences and Engineering Division under contract no. DE-AC02-05-CH11231 (van der Waals heterostructures programme, KCWF16). The electron-hole bilayer device fabrication was supported by the AFOSR award FA9550-23-1-0246. S.T acknowledges primary support from DOE-SC0020653 (materials synthesis), Applied Materials Inc., NSF CMMI 1825594, NSF DMR-1955889, NSF CMMI-1933214, NSF 1904716, NSF 1935994, NSF ECCS 2052527, DMR 2111812, and CMMI 2129412. K.W. and T.T. acknowledge support from JSPS KAKENHI (Grant Numbers 19H05790, 20H00354). R.Q. acknowledges support from the Kavli ENSI graduate student fellowship.

## Author contributions
F.W. conceived the research. R.Q. and Q.L. fabricated the devices with help from S.C., J.X and Z.C. and supervision from A.Z. and F.W. R.Q. performed the optical measurements assisted by Z.Z. J.X. and A.Y.J. R.Q. and F.W. analyzed the data with inputs from D.D.D. and L.F. Y.O. and S.T. grew $MoSe_2$, $MoS_2$, and $WSe_2$ crystals. K.W. and T.T. grew hBN crystals. All authors discussed the results and wrote the manuscript.

## Competing interests
The authors declare no competing interests.

# Figures and captions

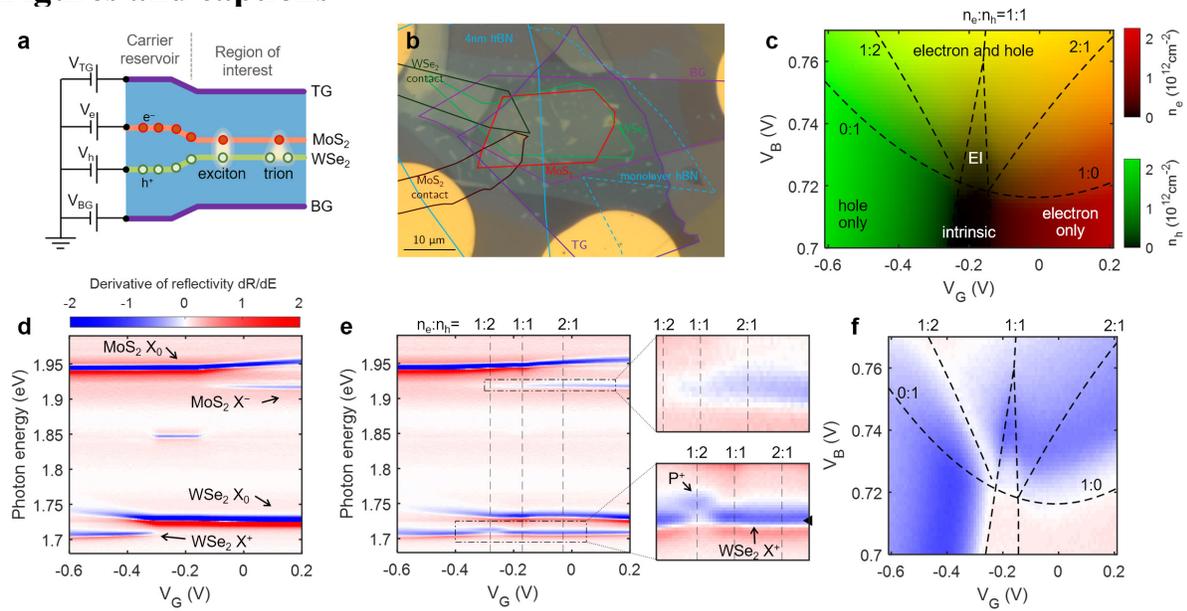

**Fig. 1 | Electrically controlled interlayer exciton and trion fluids.**
**a**, Schematic cross-section of the $MoS_2$/hBN/$WSe_2$ heterostructure device.
**b**, An optical image of $MoS_2$/monolayer hBN/$WSe_2$ heterostructure device D1, with flake boundaries outlined.
**c**, Carrier doping phase diagram of the e-h bilayer. The vertical electric field is fixed by $V_{TG} - V_{BG} = 5.6V$ while the gate voltage $V_G$ and bias voltage $V_B$ are varied. The red and green channel of the image show the electron density and the hole density respectively.
**d-e**, Energy-derivative of reflection spectrum $dR/dE$ as a function of gate voltage $V_G$. The bias voltage is fixed at $V_B = 0.68V$ (**d**) and $V_B = 0.735V$ (**e**).
**f**, $dR/dE$ at $WSe_2$ $X^+$ resonance energy (1.707eV, marked by the black triangle in **e**).

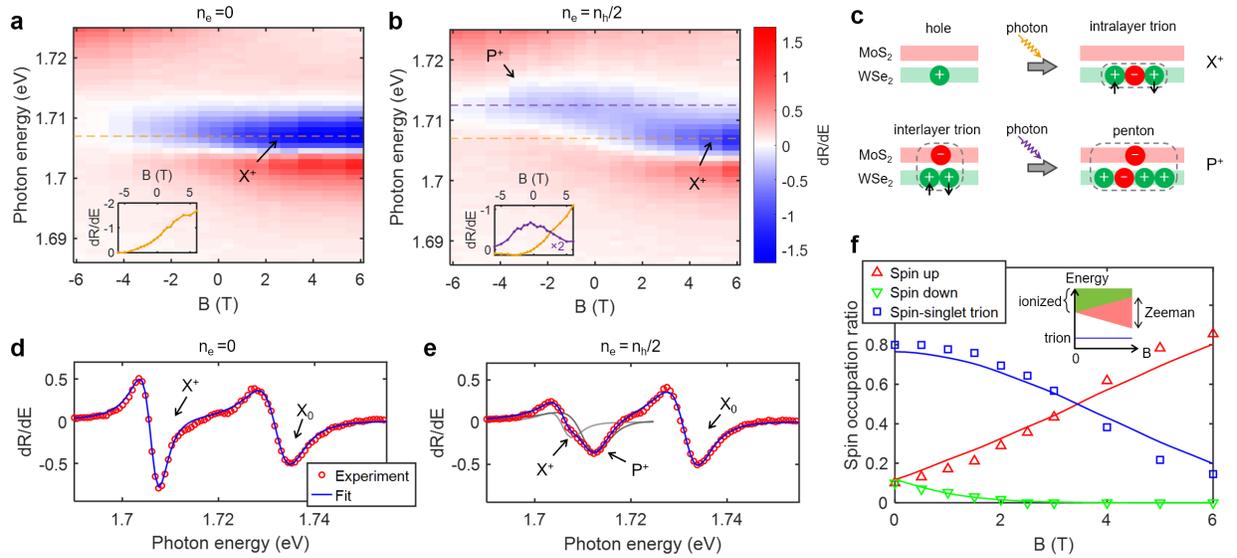

**Fig. 2 | Spectroscopic evidence of interlayer trion formation.**
**a**, d$R$/d$E$ spectrum at different perpendicular magnetic field $B$, with left-handed circularly polarized light probe. Carrier densities are $n_h = 1.0 \times 10^{12}\,\text{cm}^{-2}$ and $n_e = 0$. Inset, d$R$/d$E$ at X$^+$ peak energy as a function of $B$.
**b**, Same data as **a** but with $n_h = 1.0 \times 10^{12}\,\text{cm}^{-2}$, $n_e = 0.5 \times 10^{12}\,\text{cm}^{-2}$. Inset, d$R$/d$E$ at X$^+$ (yellow) and P$^+$ (purple) peak energy as a function of $B$. The purple line is multiplied by 2 for visibility.
**c**, Schematic illustration of resonant photon absorption process for two representative doping ratios. Red and green circles are electrons and holes respectively. Dashed gray boxes denote bound states. Black arrows represent spins.
**d-e**, Comparison of the d$R$/d$E$ spectrum between $n_e = 0$ and $n_e = n_h/2$ at zero magnetic field. The experimental spectra (open circles) are fitted to the derivative of multiple Fano peaks (solid lines).
**f**, Magnetic field dependence of spin-up, spin-down and spin-singlet trion state occupation ratio, at $n_h = 1.0 \times 10^{12}\,\text{cm}^{-2}$, $n_e = 0.5 \times 10^{12}\,\text{cm}^{-2}$. Scatters, experimental data from **b**. Solid lines, fitting results from the model described in Methods.

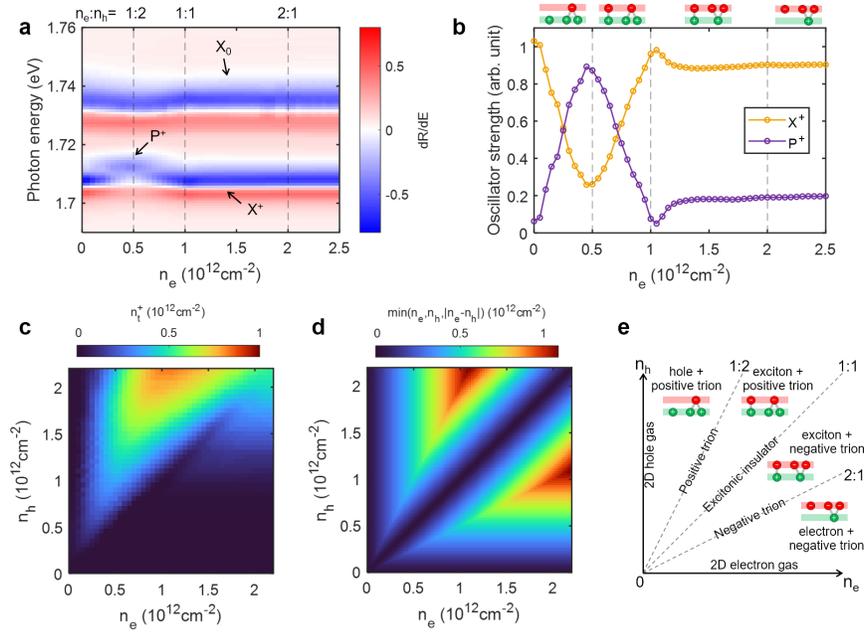

**Fig. 3 | Tunable trion-exciton-charge mixture.**
**a**, $dR/dE$ spectrum as a function of $n_e$ for fixed $n_h = 1.0 \times 10^{12}\,\text{cm}^{-2}$.
**b**, Fitted $X^+$ and $P^+$ oscillator strengths as a function of $n_e$ for fixed $n_h = 1.0 \times 10^{12}\,\text{cm}^{-2}$.
**c**, Experimentally estimated positive trion density $n_t^+$ as a function of $n_e$ and $n_h$.
**d**, Trion density from the simple physical picture described in the main text.
**e**, Schematic phase diagram of the e-h fluid.

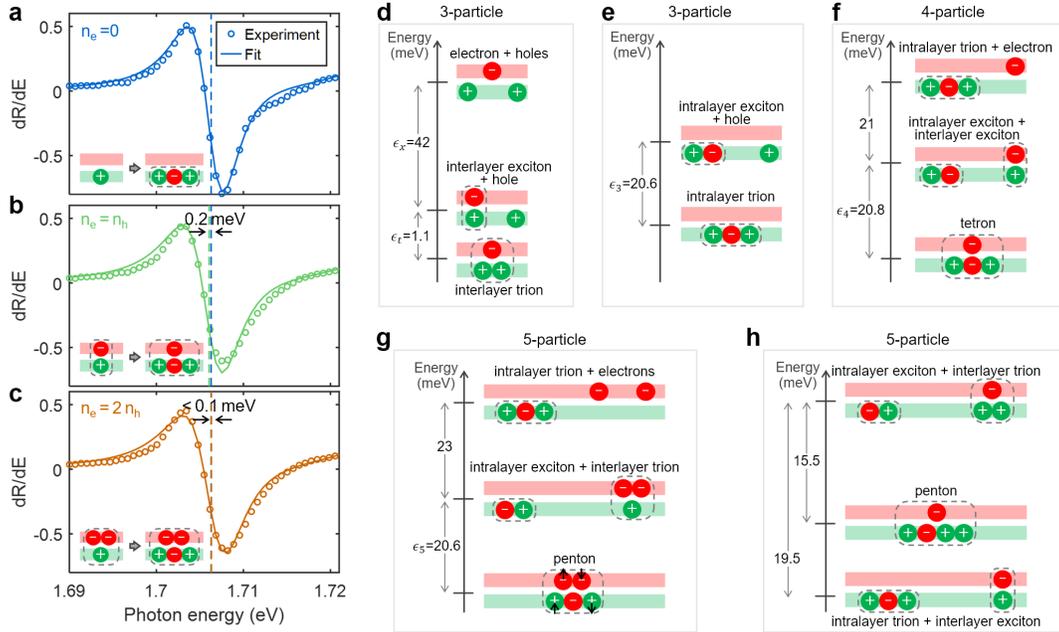

**Fig. 4 | Photoexcited high-order multiparticle charge complexes.**
**a-c,** Comparison of the X$^+$ peak for $n_e = 0, n_h$ and $2n_h$, keeping $n_h = 1.0 \times 10^{12} \text{cm}^{-2}$. Vertical dashed lines mark the fitted center energy of the Fano peak. Inset, schematic illustrations of the corresponding photon absorption process.
**d-h**, Energy diagram of multiparticle complexes with one MoS$_2$ electron and two WSe$_2$ holes (**d**), one WSe$_2$ electron and two WSe$_2$ holes (**e**), one MoS$_2$ electron, one WSe$_2$ electron and two WSe$_2$ holes (**f**), two MoS$_2$ electrons, one WSe$_2$ electron and two WSe$_2$ holes (**g**), and one MoS$_2$ electron, one WSe$_2$ electron and three WSe$_2$ holes (**h**).

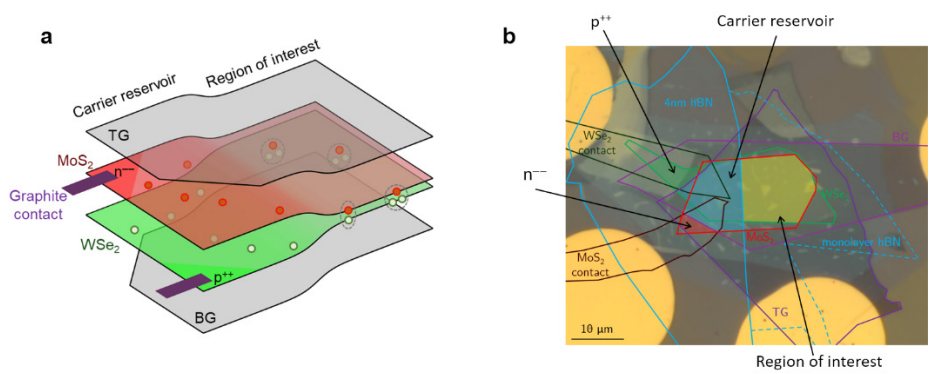

**Extended Data Fig. 1 | Detailed device structure.**
**a**, Schematic illustration of the device geometry.
**b**, Optical microscope image of device D1 with flakes outlined.

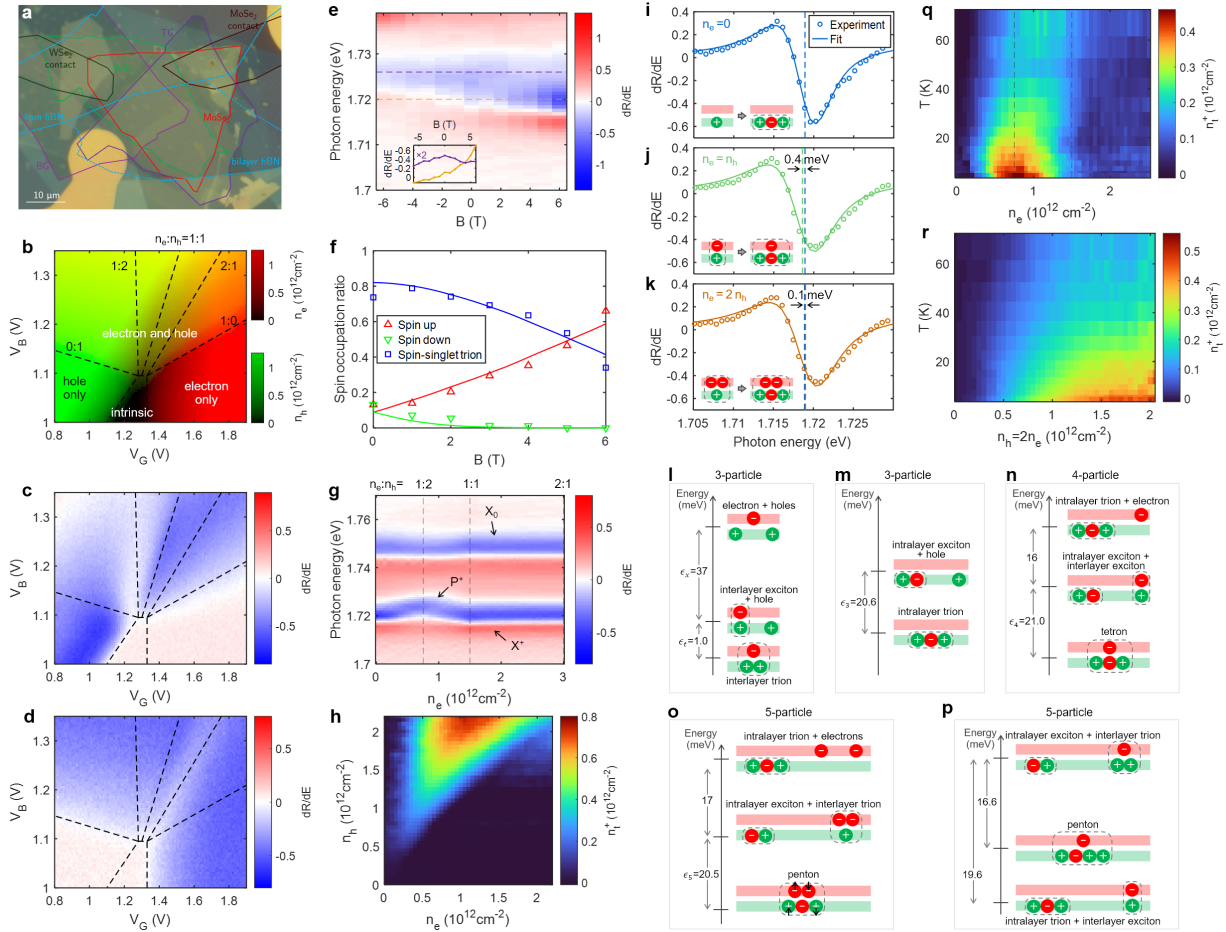

**Extended Data Fig. 2 | Additional data from MoSe$_2$/bilayer hBN/WSe$_2$ device D2.**
**a**, Optical microscope image of device D2 with flakes outlined.
**b**, Carrier doping phase diagram of the e-h bilayer.
**c**, $dR/dE$ at WSe$_2$ X$^+$ energy (1.720eV), showing a decreased X$^+$ intensity along $n_e = n_h/2$.
**d**, $dR/dE$ at MoSe$_2$ X$^-$ energy (1.623eV), showing a decreased X$^-$ intensity along $n_e = 2n_h$.
**e**, $dR/dE$ spectrum at different perpendicular magnetic field, with left-handed circularly polarized light probe. Carrier densities are $n_h = 1.5 \times 10^{12} \text{cm}^{-2}$ and $n_e = 0.75 \times 10^{12} \text{cm}^{-2}$. Inset, $dR/dE$ at X$^+$ (yellow) and P$^+$ (purple) peak energy as a function of $B$.
**f**, Magnetic field dependence of spin-up, spin-down and spin-singlet trion state occupation ratio. Scatters, experimental data from **e**. Solid lines, fitting results.
**g**, $dR/dE$ spectrum as a function of $n_e$ for fixed $n_h = 1.5 \times 10^{12} \text{cm}^{-2}$.
**h**, Experimentally estimated positive trion density $n_t^+$ as a function of $n_e$ and $n_h$.
**i-k,** Comparison of the X$^+$ peak for $n_e = 0, n_h$ and $2n_h$, keeping $n_h = 1.5 \times 10^{12} \text{cm}^{-2}$. Vertical dashed lines mark the fitted center energy of the Fano peak.
**l-p**, Energy diagram of multiparticle complexes.
**q**, Temperature dependence of the estimated positive trion density $n_t^+$ as a function of $n_e$ for fixed $n_h = 1.5 \times 10^{12} \text{cm}^{-2}$.
**r**, Temperature dependence of the estimated positive trion density $n_t^+$ along the $n_e = n_h/2$ line.

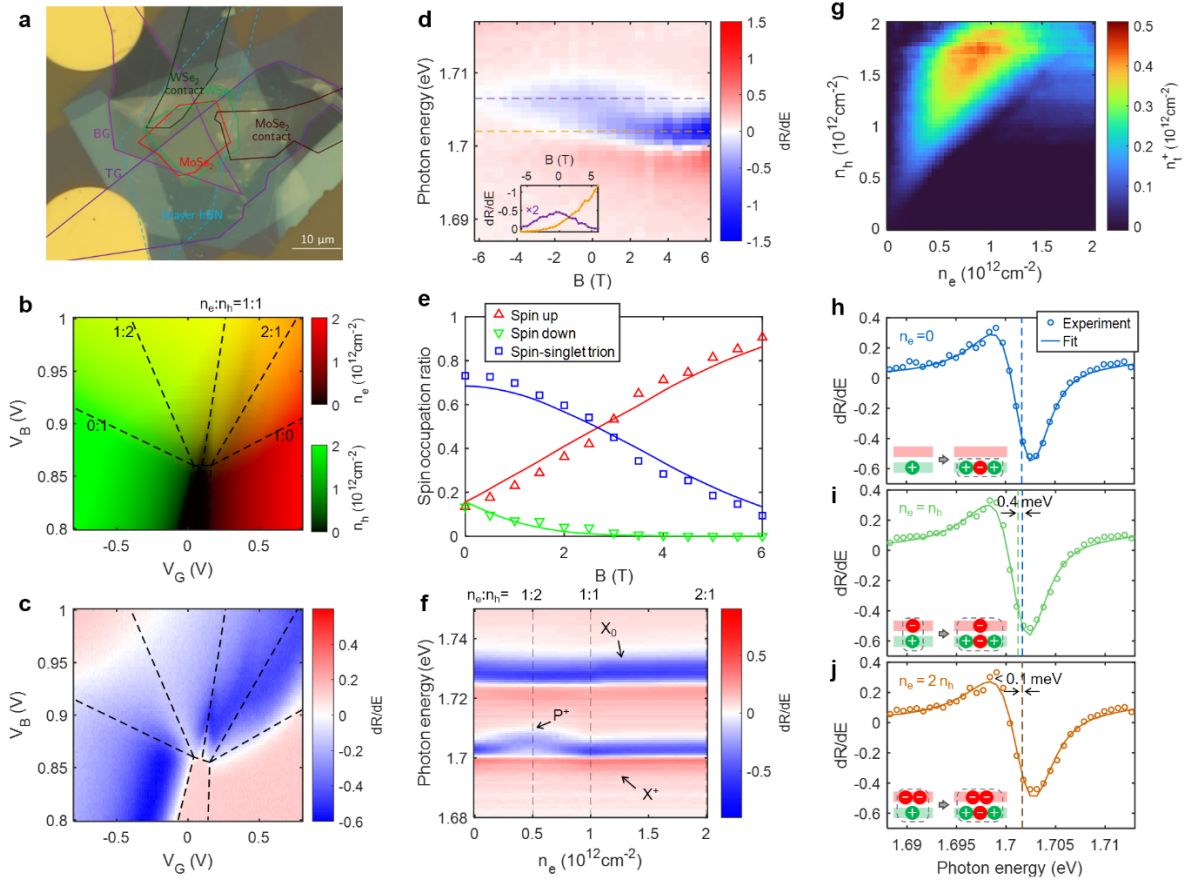

**Extended Data Fig. 3 | Additional data from MoSe₂/bilayer hBN/WSe₂ device D3.**
**a**, Optical microscope image of device D3 with flakes outlined. This device does not have a thicker (3-5 nm) hBN spacer in the contact region.
**b**, Carrier doping phase diagram of the e-h bilayer. The exciton injection in the EI phase is less efficient, leading to a distorted carrier density map near the net charge neutrality.
**c**, $dR/dE$ at WSe₂ X⁺ resonance energy (1.703eV), showing a decreased X⁺ intensity along $n_e = n_h/2$.
**d**, $dR/dE$ spectrum at different perpendicular magnetic field, with left-handed circularly polarized light probe. Carrier densities are $n_h = 1.0 \times 10^{12} \text{cm}^{-2}$ and $n_e = 0.5 \times 10^{12} \text{cm}^{-2}$. Inset, $dR/dE$ at X⁺ (yellow) and P⁺ (purple) peak energy as a function of $B$.
**e**, Magnetic field dependence of spin-up, spin-down and spin-singlet trion state occupation ratio. Scatters, experimental data from **d**. Solid lines, fitting results.
**f**, $dR/dE$ spectrum as a function of $n_e$ for fixed $n_h = 1.0 \times 10^{12} \text{cm}^{-2}$.
**g**, Estimated positive trion density $n_t^+$ as a function of $n_e$ and $n_h$.
**h-j**, Comparison of the X⁺ peak for $n_e = 0, n_h$ and $2n_h$, keeping $n_h = 1.0 \times 10^{12} \text{cm}^{-2}$.

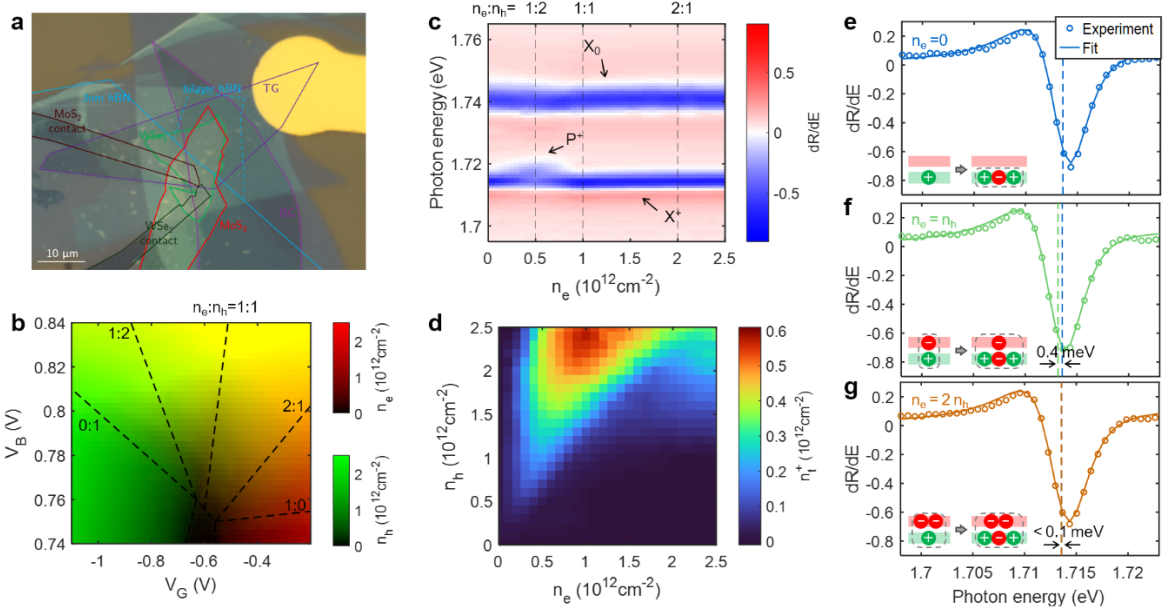

**Extended Data Fig. 4 | Additional data from MoS$_2$/bilayer hBN/WSe$_2$ device D4.**
**a**, Optical microscope image of device D4 with flakes outlined.
**b**, Carrier doping phase diagram of the e-h bilayer.
**c**, $dR/dE$ spectrum as a function of $n_e$ for fixed $n_h = 1.0 \times 10^{12} \text{cm}^{-2}$.
**d**, Estimated positive trion density $n_t^+$ as a function of $n_e$ and $n_h$.
**e-g**, Comparison of the X$^+$ peak for $n_e = 0, n_h$ and $2n_h$, keeping $n_h = 1.0 \times 10^{12} \text{cm}^{-2}$.

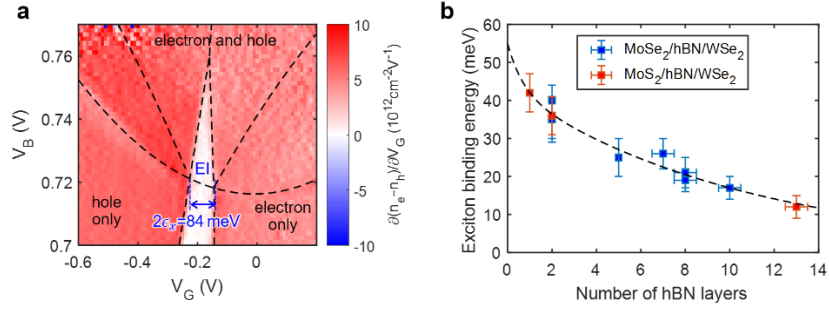

**Extended Data Fig. 5 | EI phase and interlayer exciton binding energy.**
**a**, Charge compressibility $\partial(n_e - n_h)/\partial V_G$. The single-particle gap can be determined from the width of the excitonic insulator phase at the start of exciton injection, which gives $2\epsilon_x = 84 \pm 10$ meV.
**b**, Interlayer exciton binding energy for various hBN spacer thicknesses. The dashed line is a guide to the eye. Error bars are estimated from the pixel size of the charge compressibility map.

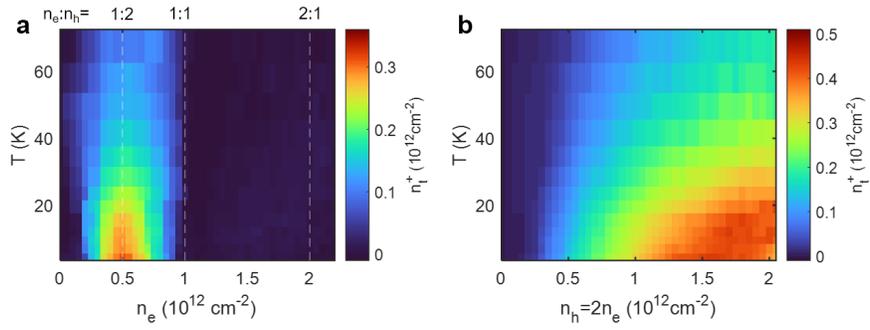

**Extended Data Fig. 6 | Temperature dependence.**
**a**, Temperature dependence of the estimated positive trion density $n_t^+$ as a function of $n_e$ for fixed $n_h = 1.0 \times 10^{12} \text{cm}^{-2}$.
**b**, Temperature dependence of the estimated positive trion density $n_t^+$ along the $n_e = n_h/2$ ratio line.

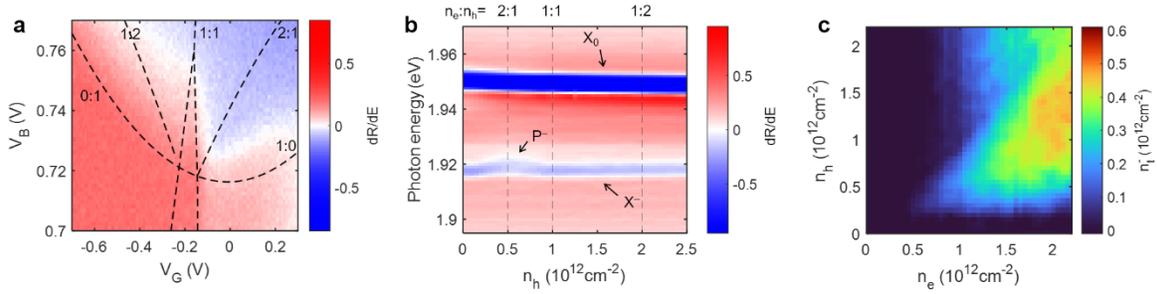

**Extended Data Fig. 7 | Spectroscopic signature of the negative trion.**
**a**, $dR/dE$ at $MoS_2$ $X^-$ resonance energy (1.922eV), showing a decrease in $X^-$ intensity centered along the $n_e = 2n_h$ line.
**b**, $dR/dE$ spectrum as a function of $n_h$ for fixed $n_e = 1.0 \times 10^{12} cm^{-2}$. The penton peak $P^-$ associated with the negative interlayer trion fluid shows up near $n_e = 2n_h$.
**c**, Experimentally estimated negative trion density $n_t^-$ as a function of $n_e$ and $n_h$, in agreement with the model shown in Fig. 3d.

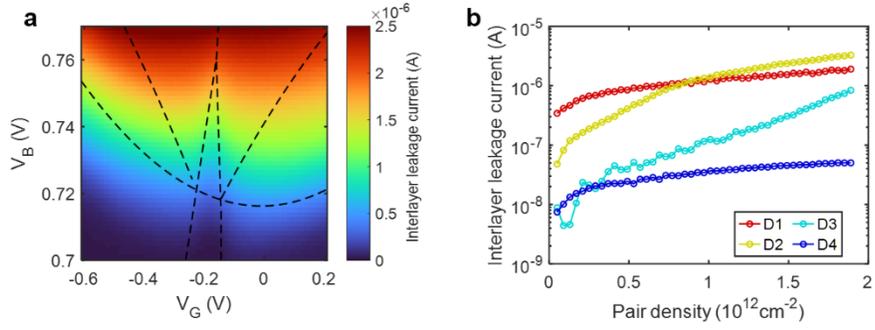

**Extended Data Fig. 8 | Interlayer tunneling current.**
a, Interlayer leakage current for device D1 as a function of $V_G$ and $V_B$.
b, Interlayer current as a function of pair density at net charge neutrality for four devices D1-D4. The average estimated interlayer recombination lifetime is 0.2μs, 0.3μs, 0.8μs and 2.4μs for the four devices respectively.